ORIGINAL ARTICLE

# A clinical update on Antibiotic Resistance Gram-negative bacteria in Malaysia- a review

**Fazlul MKK[1], Shah Samiur Rashid[1], Nazmul MHM[2], Zaidul I.S.M[3], Roesnita Baharudin[4], Aizi Nor Mazila RAMLI[1]**

[1]Faculty of Industrial Sciences and Technology, Universiti Malaysia Pahang, Gambang, 26300 Pahang, Malaysia

[2]Graduate School of Medicine, Perdana University, Jalan MAEPS Perdana, Serdang, 43400 Selangor, Malaysia

[3]Department of Pharmaceutical Technology, Kulliyyah of Pharmacy, International Islamic University, Kuantan Campus, 25200 Kuantan, Pahang, Malaysia

[4]Department of Pathology, Hospital Tengku Ampuan Afzan, 25100 Kuantan, Pahang, Malaysia

## Abstract

Antibiotics are the wonder discoveries to combat microbes. For decades, multiple varieties of antibiotics have been used for therapeutic purposes in hospital settings and communities throughout the world. Unfortunately, bacteria have become resistant to commonly prescribed antibiotics. This review aims to explore the development, challenges, and the current state of antibiotic resistance available literature in Malaysia. Recently published data from Pubmed, Scopus, Google scholar's and other internet source publication on the antibiotic resistance of gram-negative bacteria were further reviewed and analysed in this study. This review reiterates the antibiotic resistance among the gram-negative bacteria is increasing and they are becoming resistant to nearly all groups of antibiotics. The antibiotic treatments are minimal and hard to treat in multi-drug resistance bacterial infection, resulting in morbidity and mortality. The prevalence rate of antibiotic resistance from the literature suggests that educating patients and the public is essential to prevent and control the spread of antibiotic resistance. In particular, there is an urgent need for a surveillance system of regular monitoring on the microbiomes, the discovery of novel antibiotics and therapeutic application of antibiotics are mandatory.

## 1. INTRODUCTION

Antimicrobial resistance creates a significant health problem worldwide by its various factors, especially in hospital settings. The primary factors include the exposures to cross infection and broad-spectrum antimicrobials. Understanding the extent and type of antimicrobial use might help in developing an awareness of antibiotic resistance. Becoming resistant to multiple antibiotics is a serious threat of gram-negative bacteria through out the world. These resistant bacteria most of the times prolong the treatment procedures and also transfer its' genetic material to other bacteria to transform the bacteria into a drug-resistant strain (World Health Organization, 2017, 27 February). The occurrences of multidrug-resistant (MDR) pathogens are contineously increasing although many different ways are employed to combat the antibiotic resistance of the pathogens (Laxminarayan et al., 2013). Millions of lives are being saved by transforming antibiotics into medicine while prevalence of antibiotic resistance is reported worldwide (Golkar, Bagasra, & Pace, 2014). Antibiotics play essential roles in achieving significant advances in medicine and surgery through saving millions of patients' lives. Many decades after the first patient was treated with antibiotics yet bacterial infections are a major threat to health managements (Spellberg & Gilbert, 2014). Inappropriate use of antibiotic therapy associated with disease can lead to the development of drug-resistant strains (Deris, 2015).

Correspondence Author: Dr Shah Samiur Rashid
Email address: samiur@ump.edu.my

The antibiotic resistance crisis is due to the overuse and misuse, along with incapability to develop new drug by the pharmaceutical companies. The Centres for Disease Control and Prevention (CDC) has classified some bacteria as presenting urgent, dangerous, and concerning threats. Antibiotic resistance gram-negative bacteria including extended spectrum β-lactamases (ESBLs) and multidrug-resistant have emerged and spread throughout Malaysia.

## 2. METHODS

### 2.1 Literature search, definition and selection strategy

In this article, it was reviewed and summarized the current critical knowledge on the epidemiolocal as well as molecular characterization of antibiotic resistance in Malaysia. The National Surveillance of Antibiotic Resistance (NSAR) report by the Ministry of Health, Malaysia have revealed data until 2017 from 39 hospital-based microbiology laboratories which included thirty-seven microbiology laboratories of government hospitals, and 2 hospital based university. The published articles in Pubmed, Scopus, Google scholar's and other recent internet source publication on the antibiotic resistance of gram-negative bacteria were further reviewed and included in this study.

## 3. Result and Discussion

The most critical group of gram-negative bacteria which causes diseases and become resistant to multiple antibiotics was highlighted. The multidrug resistant bacteria such as *Acinetobacter*, *Pseudomonas* and various Enterobacteriaceae including *Klebsiella*, and *E. coli* can cause severe and often deadly infections those are frequently reported in the hospitals and clinics. However, many studies reported that gram-negative bacteria showed multiple-antibiotic resistance due to misuse of antibiotics, longer duration of prior antimicrobial therapy and a higher number of antimicrobial applications to control infections. Highest and medium priority categories of bacteria those cause bloodstream infections and pneumonia was reviewed under current scope. This study would give a clear view on the urgency (critical, high and medium priority) of new antibiotics development. It was also presented the most dominant bacteria causing various types of diseases throughout the world as well as in Malaysia.

***Acinetobacter***

*Acinetobacter* is a complex genus with the existence of multiple species. The species commonly cause nosocomial infections, pneumonia, catheter-associated bacteremia, soft tissue and urinary tract infections (Wong et al., 2017). *Acinetobacter* spp. described as natural transformable bacteria and can become resistant to many different classes of antibiotics (Perez et al., 2007).

In 2014, fifty-four *Acinetobacter* isolates (72.2%) obtained from a tertiary hospital in Terengganu, Malaysia. This study revealed the resistance rate of tetracycline (87%), piperacillin-tazobactam (72.2%), cefotaxime (72.2%), ceftazidime (72.2%), cefepime (72.2%), levofloxacin (70.4%), ampicillin-sulbactam (68.5%). In addition, gentamicin (66.7%), ciprofloxacin (66.7%), tobramycin (64.8%), doxycycline (61.1%) and amikacin (57.4%). The isolates also showed high resistance to carbapenems as 77.8% of the isolates found to be resistant to meropenem and 74.1% to imipenem. The MIC values of carbapenem-resistant isolates was >32 $\mu$g/mL. 39 (72.2%) isolates out of 54 strains were multidrug-resistant (MDR) means resistant to more than three classes of antibiotics against bacteria. All MDR isolates were resistant to meropenem and imipenem. Polymyxin resistance which was determining the MIC values for polymyxin B. of the 54 isolates. The resistant was 25.9% to polymyxin (MIC $\geq$ 4 $\mu$g/mL, and four isolates had MIC values of >128 $\mu$g/mL for polymyxin B. These 14 polymyxin-resistant isolates classified as extensive drug resistant (XDR) isolates (Lean et al., 2014). Moreover, a molecular study on 40 isolates of *Acinetobacter* spp.were resistant to ciprofloxacin (55%), imipenem (67.5%), amikacin (50%), gentamicin (52.5%), ampicillin (75%), ceftazidime (60%), meropenem (92.5%), piperacillin (77.5%), cefoperazone (60%) and tazobactam/piperacillin (67.5%) respectively. Meropenem showed the maximum resistance (92.5%) followed by piperacillin (77.5%) and ampicillin (75%). In addition, 5% *Acinetobacter* strains were resistant to one antibiotic, 10% strains to two antibiotics, and 85% strains to three or more antibiotics (Nazmul, Jamal, & Fazlul, 2012). National Surveillance of Antibiotic Resistance, Malaysia have reported that increase resistance rates of various types antibiotics from 2013 to 2017. Studies showed that ampicillin/sulbactam, ceftazidime, imipenem, meropenem, gentamicin, amikacin, ciprofloxacin, cefoperazone/sulbactam, and piperacillin/tazobactam were resistance to *Acinetobacter* spp. The resistance rate of ampicillin/sulbactam noted to slightly increase to 53.8% in 2015 compared to 53.3 in 2014. The highest resistance was observed for meropenem 58.8% and the least resistance were detected for colistin 0.8% while in 2016 Meropenem resistance increased to 60.8% (M. o. H. Institute for Medical Research, Malaysia, 2017, 15th September).

In 2016, antibiotic susceptibility test by S.Y. Ng observed 15 out of 22 isolates (68.18%) categorized as multidrug resistant organisms (MDROs) due to resistant by one agent in three or more antibiotic. The significant strains were resistant towards cephalexin (95.45%), followed by enrofloxacin (59.09%), amoxicillin-clavulanic acid (54.55%), sulphamethoxazole-trimethoprim (54.55%) and tetracycline (50%) respectively. Five out of six *A. Baumannii* isolates (83.33%) were classified as MDR by antibiotic susceptibility test, and all (6) *Acinetobacter baumannii* isolates were resistance to all antibiotics. From the antibiotic susceptibility profile, all the six *A. Baumannii* isolates (100%) were resistant to cephalexin. Among the six isolates, one strain was resistant to tetracycline and enrofloxacin, two strains were resistant to sulphamethoxazole-trimethoprim, and rest three isolates were resistant to amoxicillin-clavulanic acid (Ng, Abdul Aziz, Bejo, & Dhaliwal, 2016). In 2007, a mixed growth of imipenem-resistant strain of *K. pneumonia*, *Acinetobacter* spp. and a *Proteus* spp. was observed at University of Malaya Medical Centre (UMMC), Kuala Lumpur, Malaysia. The *Acinetobacter* spp. was resistant to imipenem, ceftazidime, cefuroxime, cefotaxime, cefepime, and meropenem respectively while sensitive to a number of antibiotics (amikacin, gentamicin, netilmicin, ciprofloxacin, cefoperazone–sulbactam and ampicillin-sulbactam) (Palasubramaniam, Karunakaran, Gin, Muniandy, & Parasakthi, 2007).

The presence of multidrug-resistance (MDR) in *Acinetobacter baumannii* strains was observed in a tertiary-care center, Subang Jaya, Selangor, Malaysia by Jayakayatri in 2016. These studies revealed that the strains of *Acinetobacter baumannii* were resistance to 11 types of antimicrobials (polymyxin B, ampicillin/sulbactam, ceftazidime, meropenem, imipenem, amikacin, gentamicin, ciprofloxacin, levofloxacin, trimethoprim/sulfamethoxazole, and tigecycline). A total number of 53% strains were classified as multidrug-resistance (resistant to 4 or more antibiotics) (Desa, Yong, Navaratnam, Palanisamy, & Wong, 2016). The accumulated study data on *Acinetobacter* can be beneficial for proper antibiotic treatments and infections control management. According to the World Health Organization (WHO), carbapenem-resistant for *Acinetobacter baumanni* are in critical situations (World Health Organization, 2017, 27 February). Over the past decades, different types of antibiotics including meropenem, tetracycline, ceftazidime, and cefuroxime have shown the highest resistance to *Acinetobacter* species.

Meropenem shows the highest frequency of resistance among all used antibiotics. The highest antibiotic resistance rate of *Acinetobacter* spp. has shown in table 1 from the accumulated data observed in Malaysia. An increased in resistant rate was noted for imipenem (61.7%), meropenem (61%), gentamicin (53.2%), amikacin and (49.8%), in 2017 compared to 2010. An increasing trend of antibiotic resistance was observed among the clinical samples of *Acinetobacter* spp. between 2010 and 2017. Among the 5 used antibiotics, imipenem 61.7% indicated the highest resistant in 2017 while 54.8% in 2010. The trends of antibiotic resistance are increasing among the used antibiotics past 7 years (2010 to 2017) have shown in figure 1. Although The lowest resistance was observed for gentamicin (53.2%) in 2017 but 48% in 2010 (M. o. H. Institute for Medical Research, Malaysia, 2017, 15th September).

***Pseudomonas aeruginosa***

*Pseudomonas aeruginosa* commonly caused different types of disease including community-acquired and nosocomial-acquired pneumonia. Over the years, the development of *Pseudomonas aeruginosa* resistance is increasing globally due to the overuse of antibiotics (Yayan, Ghebremedhin, & Rasche, 2015). The resistance of *P. aeruginosa* occurs after 3–4 days of the initiation of treatment (Juayang et al., 2017). The most severe nosocomial infection correlates with increased resistance to antibiotics with high morbidity and mortality (Sonmezer et al., 2016).

In 2007, a multi-resistant strain of *Pseudomonas aeruginosa* was sensitive to polymyxin B, intermediate to amikacin and resistant to netilmicin, piperacillin, piperacillin-tazobactam, imipenem, ceftazidime, cefoperazone, ciprofloxacin, gentamicin, and meropenem (Palasubramaniam et al., 2007). In addition, fifty-four clinical isolates of *Pseudomonas aeruginosa* from Selayang Hospital, Malaysia has revealed the rates of antimicrobial resistance. The strains were resistant 9.25 % to ciprofloxacin, meropenem, and cefepime respectively. The resistance rates were observed for amikacin 12.97%, tazobactam 27.78%, imipenem 16.67%, cefoperazone 27.78%, ceftazidime 29.63%, gentamicin 14.82% and 50% to piperacillin (Fazlul, Zaini, Rashid, & Nazmul, 2011). Moreover, a study on *P. aeruginosa* at tertiary care hospital in Malaysia by Raja et al., described the rates of antimicrobial resistance of 12.9% gentamicin, 10.1% netilmicin, 10.9% ceftazidime, 11.3% ciprofloxacin, 9.9% imipenem, 10.8% piperacillin, 9.4% piperacillin-tazobactam, 6.73% amikacin and 0% to polymyxin B while 5.74% of the strains were found to be (MRD) multidrug-resistant (Raja & Singh, 2007).

*Pseudomonas aeruginosa*, the common pathogen causes nosocomial infections. In 2009, a molecular characterization was determined for 48 clinical isolates of *P. aeruginosa* from 6 public hospitals in Malaysia. The obtained data was analyzed by antimicrobial susceptibility test and DNA fingerprinting techniques. In these study *P. aeruginosa* isolates were resistant to tetracycline (73%), chloramphenicol (60%), cefotaxime (40%), ceftriaxone (31%), cefoperazone (29%), ticarcillin (25%), piperacillin (23%), and imipenem (21%). Less than 20% of *P. aeruginosa* were sensitive to ceftazidime, gentamicin, cefepime, ciprofloxacin, amikacin, piperacillin-tazobactam, and aztreonam. Moreover, 33 strains of *P. aeruginosa* (68.75 %) were multidrug resistant, and aztreonam was the most effective antibiotic (Lim et al., 2009). In a similar study, ninety (90) randomly selected non-replicate imipenem-resistant *P. aeruginosa* isolates were collected from UMMC, a university teaching hospital at Kuala Lumpur, Malaysia. All these clinical specimens were from nonrepetitive patients (57 females and 33 males). These study reported that antibiotic resistance profiles of *P. aeruginosa* isolate by using disk-diffusion tests. Ceftazidime 63% showed the highest resistant followed by amikacin 62% and least resistant to netilmicin 48% (Khosravi, Tee Tay, & Vadivelu, 2010) in *P. aeruginosa*. National Surveillance of Antimicrobial Resistance (NSAR), 2017 reported that the imipenem 7.8% showed the highest resistance in 2017 while 9.1% in 2016. Among the used antibiotics for *P. aeruginosa*, only gentamicin has increased in resistance pattern from 3.1% in 2016 to 5.6% in 2017. The least resistance to colistin 0.3% was observed in 2015 while 2.7% in 2013 (M. o. H. Institute for Medical Research, Malaysia,, 2016). Different range of resistance rate of amikacin (3.5%), ceftazidime (6.9%), cefepime (5.4%) and meropenem (6.6%) was observed in 2017 compared to 2017 (M. o. H. Institute for Medical Research, Malaysia, 2017, 15th September). According to the World Health Organization (WHO), carbapenem-resistant for *Pseudomonas aeruginosa* is in critical situations (World Health Organization, 2017, 27 February).

Among the different types of antibiotics, ceftazidime showed the higest resistance to *P.aeruginosa* followed by piperacillin and least resistance by colistin between 2002 and 2017. The variation of antibiotic resistance rate was observed among all used antibiotics since 2002 to 2017 (Figure 2). Meropenem (8.1%), ceftazidime (8.9%), and cefepime (5.5%) showed the decreasing trends in 2016 compared to resistance in 2015 (M. o. H. Institute for Medical Research, Malaysia, 2017, 21st September,).

### *Klebsiella* species

*Klebsiella* species causes nosocomial and community-acquired infections worldwide (Moradigaravand, Martin, Peacock, & Parkhill, 2017). Severe lung and bloodstream infections which are challenging to treat due to its multidrug resistance properties of *Klebsiella pneumoniae* (Jana et al., 2017).

The prevalence and antibiotic susceptibility of Gram-negative bacteria among infants admitted to the neonatal intensive care unit (NICU) in a tertiary teaching hospital in Malaysia described a total of 34 stool specimens obtained from infants of multidrug-resistant *k. pneumoniae* unusually high resistance to polymyxins. Three colistin resistant *k. pneumoniae* have also detected by E-test analysis (Low et al., 2017). Thirty-nine isolates of *k. pneumoniae* were resistant to more than three antimicrobial classes which classified as multidrug-resistant (MDR). Ampicillin-resistance was highest in *Klebsiella* spp. (84.3%) (Nor et al., 2015) described out of 721 gram-negative isolates and 21.2% was *Klebsiella* spp.

A recent study of 93 Malaysian patients who was affected by the multidrug-resistant of *K. Pneumonia* isolates at University of Malaya Medical Center,

Kuala Lumpur, Malaysia. These reports demonstrated higher resistance rates of ampicillin (100%) followed by aztreonam (98.9%) and cephalosporins including ceftriaxone (100%), cefoperazone (100%), cefotaxime (100%) showing the maximum resistance followed by cefuroxime (98.9%) and ceftazidime (97.8%) against *K. Pneumonia*. Also, the isolates were susceptible to cefoxitin, except three (3.2%) isolates (strain K106, M40, and NDM-2012). The resistance rate of piperacillin-tazobactam (43%) was lower then other β-lactam/β-lactamase inhibitor combinations (94.6% for amoxicillin-clavulanate and ampicillin-sulbactam 98.9%). Non-susceptibility to trimethoprim/sulfamethoxazole, gentamicin and amikacin were observed in 90.3%, 74% and 5.4% of the isolates respectively (Al-Marzooq, Yusof, & Tay, 2015). In addition, imipenem-resistant strain of *K. pneumonia* in 2007 by Selvi at UMMC, Kuala Lumpur Malaysia confirmed resistant to the extended-spectrum β-lactams and sensitive to gentamicin, amikacin, and netilmicin (Palasubramaniam et al., 2007).

In 2017, carbapenem-resistant of *K. pneumoniae* were resistant to tetracycline except two strains were intermediate, but only one strain was resistant to tigecycline in Malaysia. Among all the isolates, 47.06% strains were resistant to cephalosporin groups of antibiotics while all bacterial strains were sensitive to colistin and resistance to atleast one carbapenem (MIC = 4 - >32 μg/ml) antibiotic. Resistance to imipenem significantly associated with in-hospital mortality (P = 0.043). Two strains (K/1311-35 and K/1304-16) were resistant to imipenem (MIC = 4 – 6 μg/ml) while sensitive to meropenem and ertapenem. One isolates of *K. pneumoniae* (K/1312-3) was sensitive to imipenem but resistant to meropenem and ertapenem (MIC = 6 – 16 μg/ml). Again, three strains (17.65%) were resistant to aminoglycoside group of antibiotics (MIC = 24 - >256 μg/ml for gentamicin and tobramycin; MIC = 64 - >256 μg/ml for amikacin) while thirteen strains (76.47%) were resistant to all types of fluoroquinolones antibiotics (MIC = 8 - >32 μg/ml) (Low et al., 2017).

*Klebsiella pneumonia* showed an increasing trend of antibiotic resistance over the years. National surveillance of antibiotic resistance reported that the resistant to carbapenems, namely imipenem (2.4%) and meropenem (2.8%) increased in 2015 compared to 2014 while cefuroxime showed higher resistance (28.4%) between 2013 and 2017. The least resistance was observed by imipenem 1.3% in 2013 past 5 years. The resistance rates of aminoglycosides namely gentamicin and amikacin remain reasonably stable over the years since 2013 to 2015. The resistant rate of carbapenems (imipenem 2.7% and meropenem 2.9%) were increased in 2017 compared to while imipenem 2.3% and meropenem 2.6% in 2016 (M. o. H. Institute for Medical Research, Malaysia,, 2016). The emergence of 321 carbapenem-resistant *Klebsiella pneumonia* (CRKP) were resistant to 2$^{nd}$ and 3$^{rd}$ generation cephalosporin tested included cefuroxime, cefepime, cefotaxime, ceftazidime, cefoperazone, trimethoprim-sulfamethoxazole, and amoxicillin-clavulanate in 2015. A few CRKP isolates of 5/13 (38.5%) were susceptible to aminoglycosides groups (gentamicin, amikacin, and netilmicin). Meanwhile, 4/13 (30.7%) isolates were susceptible to ciprofloxacin and 4/30 (13.3%) to piperacillin–clavulanate. Eight isolates (61.5%) were resistant to all tested antibiotics except colistin and tigecycline (Hamzan, Yean, Rahman, Hasan, & Rahman, 2015).

Among all the recent studies on *Klebsiella pneumonia* have shown high resistance impact on the different antibiotics namely polymixins, ampicillin, extended-spectrum β-lactams, tetracycline and ampicillin/sulbactam whereas the least resistance to imipenem, meropenem, and amikacin.

***Escherichia coli***

*Escherichia coli* is the most common agent of various types of infection especially urinary tract infections (UTI), cholecystitis, cholangitis, diarrhoea, neonatal meningitis and pneumonia etc. The leading cause of bacteremia is due to Gram-negative *Escherichia coli* (Woodford, Turton, & Livermore, 2011).

Noor Safina included a total of 721 cases to identify the resistance rates of gram-negative bacteria in 2015. Among all the patients 54% were male, and 46% were female. The 3 most common organisms isolated in the total population, *Escherichia coli* (41.6%) were the most prevalent organisms in both the prophylaxis (60.7%) and no prophylaxis (39.8%). Among the antibiotic resistance, *E.coli* showed a higher resistance to ampicillin (67.7%), cefuroxime (15.3%) and gentamicin (7.3%) respectively (Nor et al., 2015).

In 2009, a study of Lim and his colleagues examined 47 nonrepeat *E. coli* isolates collected in 2004 from 47 different patients of 5 public hospitals located in different location of Malaysia. The participating hospitals were Kota Bharu Hospital ($n$ = 10), Sultanah Aminah Hospital ($n$ = 23), Kuala Lumpur Hospital ($n$ = 10), Ipoh Hospital ($n$ = 2), and Queen Elizabeth Hospital ($n$ = 2). The data indicated a high prevalence of resistance to ampicillin (77%), piperacillin (64%), tetracycline (53%), trimethoprim-sulfamethoxazole (43%), cefoperazone, kanamycin (30%), Nalidixic acid (28%), chloramphenicol (26%), ciprofloxacin (23%), gentamicin (21%), amoxicillin-clavulanic acid (17%), ceftriaxone ceftazidime, and aztreonam 11% each, and amikacin 2% among all the *E. coli* isolates (Lim et al., 2009). Among them, 36 isolates (76.5%) were multidrug-resistant while cephalosporin resistant isolates were also resistant to ampicillin and 71% of them resistant to tetracycline. A large number of strains (72%) were resistant to aminoglycosides (Lim et al., 2009). Meanwhile, out of 76 gram-negative isolates of *E. coli* were resistant to amoxicillin/clavulanate (30.8%, $n$ = 16), ceftriaxone (42.3%, $n$ = 22), ceftazidime (28.8%, $n$ = 15), cefoxitin (28.8%, $n$ = 15), aztreonam (36.5%, $n$ = 19), and polymyxin B (23.1%, $n$ = 12) in hospitalized preterm infants. Among all the strains, 20 isolates of *E. coli* were resistant to more than three antibiotics categorized as multidrug-resistant (MDR) (Yap et al., 2016). In another study in 2015, ampicillin showed the highest resistance to 67% compared to 68.9% in 2013. The least resistance by meropenem (0.5%) in 2015 had slightly increased to 0.9% in 2016 and reduced to 0.7% in 2017. In blood samples, cefuroxime and cefotaxime resistance rate rose to 25.4% and 25.9% in 2015 respectively. According to National surveillance of antibiotic resistance, Malaysia, the resistance rate of cephalosporins (cefuroxime 25.9% and cefotaxime 25.4%) has slightly increased in 2017 compared to 23.5% and 23.4% respectively in 2016. Even though imipenem (0.8%) and meropenem (0.7%) resistance rate were low in 2017, but an increasing trend of resistance was observed between 2010 and 2017 (Figure 4).

Among the tested antibiotics, cefuroxime (25.9%) showed the highest resistant to *E. coli* while least resistance by meropenem (0.7%) in 2017. Moreover, cefotaxime (25.4%), ciprofloxacin (24.1%) and ceftazidime (17.7%) were the highest resistance to *E. coli* from 2013 to 2017. The resistance trends of antibiotic has increased as well as decreased against *E.coli* from 2010 to 2017 ( Figure 3) (M. o. H. Institute for Medical Research, Malaysia, 2017, 21st September,).

Since 2009 to 2017, it was reported about *E. coli* resistance against various antibiotics. Among all the used antibiotics, ampicillin showed the highest resistant (77%) followed by ceftriaxone (42.30%) and cefuroxime (25.9%). The least resistance was observed by meropenem 0.2% between 2009 and 2017 (Table 3).

The highest resistance trends were shown among the different types of the antibiotic to bacteria (*E. coli*, *Acinetobacter* spp., *Klebsiella* spp., *P. aeruginosa*) last 5 years (2013 to 2017) based on National surveillance of antibiotic resistance, Malaysia, 2017 (Figure 5). Among all the tested antibiotics, meropenem have shown the highest resistance rate of 58.6%, 63.2%, 60.6% and 61.1% to *Acinetobacter* spp. while lowest resistance rate of 0.3%, 0.2%, 0.5%, 0.9% and 0.7% to *E. coli* between 2013 to 2017. Moreover, imipenem (61.1%), ceftazidime (58%), ampicillin/sulbactam (57.1%), gentamicin (53.2%) and amikacin (49.8%) showed the highest resistance to *Acinetobacter* spp. past 5 years (2013 to 2017). In 2017, Meropenem (0.7%) were least resistance to *E.coli* while amikacin (3.5%) to *Pseudomonas aeruginosa*, polymixin B (0%) to *Acinetobacter* spp. and imipenem (2.7%) to Klebsiella pneumoniae (M. o. H. Institute for Medical Research, Malaysia, 2017, 15th September).

The antibiotic resistance trends among the *Klebsiella* spp., imipenem (80%) showed the highest increasing resistance trends followed by meropenem (71%), amikacin (53%), cefepime (40%), and amoxicillin/clavulanic acid (21%). In addition, gentamicin showed a decrease resistance trend by 19% followed by ciprofloxacin 1% between 2013 and 2017. In *E. coli.* isolates imipenem (167%) showed the highest increasing resistance trends followed by meropenem (133%), amikacin (88%), cefepime (32%), gentamicin (27%) and amoxicillin/clavulanic acid (21%) between 2013 and 2017. Among the most commonly prescribed antibiotics, ceftazidime (106%) showed the highest increasing resistance trends followed by cefoperazone/ sulbactam (18%), amikacin (14%), imipenem (13%), and gentamicin (11%) to *Acinetobacter* spp. between 2013 and 2017. In addition, polymyxin B showed a decrease resistance trend by 100%. In *P. aeruginosa*, amikacin showed an increasing resistance trend by 21% while cefepime was 8% since 2013 to 2017. Rest of the antibiotics (Meropenem 20%, ceftazidime 14%, and gentamicin 3%) showed a decrease resistance trend between 2013 and 2017 (Figure 5).

### *Haemophilus influenzae*

*Haemophilus influenzae* is an indigenous bacterium of the upper respiratory tract infection and also with *Streptococcus pneumonia* can cause respiratory tract infections (Seyama et al., 2016). The highest resistance rate was observed for trimethoprim/ sulfamethoxazole (41.4 %) to *Haemophilus influenzae* in 2015. The least resistance to cefotaxime (2.9 %) in 2015 has decreased compared to 5.3% in 2013. A report by NSAR, Malaysia revealed that ampicillin resistance increased to 25.7% in 2015 compared to 23.2% in 2014 while cefuroxime rose to 5.5% in 2015 compared to 2.5% in 2014 and trimethoprim/sulfamethoxazole decreased to 41.4% in 2015 compared to 45.5% in 2014 (M. o. H. Institute for Medical Research, Malaysia,, 2016).

### *Salmonella*

*Salmonella* spp. is a well-known pathogen causes different types of infection by consumption of contaminated food that leads to thousands of deaths per year globally (Calayag, Paclibare, Santos, Bautista, & Rivera, 2017). National surveillance of antibiotic resistance, Malaysia reported that there was a reduction in the resistance rates to ampicillin (6%), chloramphenicol (1.5%), ceftriaxone (0.3%), and trimethoprim/sulfamethoxazole (5.3%) in 2015 compared to 2014. In addition, resistance to ceftriaxone was 0% in 2013. Ciprofloxacin resistance was observed to be markedly reduced to 4.6% in 2017 compared to 8.3% in 2016. Ciprofloxacin showed a decreasing trend of resistance (45%) between 2013 and 2017. In 2017, Chloramphenicol, Ciprofloxacin, and Co-trimoxazole showed a resistance rate of 2.2 %, 4.6% and 5.4% respectively (M. o. H. Institute for Medical Research, Malaysia, 2017, 15th September). Moreover, a Recent study from clinical isolates of Selayang hospital in Malaysia reported that nontyphoidal *Salmonella* (NTS) isolates demonstrated 18.4%, 10.5% and 2.6% resistance towards ampicillin, trimethoprim-sulfamethoxazole, and ciprofloxacin respectively (Abu et al., 2016).

Among the recent studies in Malaysia, nontyphoidal *Salmonella* (NTS) isolates were resistance to ampicillin (18.4%) while 13% resistance in 2017 (M. o. H. Institute for Medical Research, Malaysia, 2017, 15th September). This ampicillin resistance of 18.4% could be an alarming issue and need further monitoring.

### *Neisseria gonorrhea*

Gonorrhea is a sexually transmitted infection (STI) caused by variable strains *Neisseria gonorrhoeae,* and worldwide it is a worrisom matters in public health priority (Wi et al., 2017). A study of NSRA reported that Penicillin resistance has slightly reduced to 51.6% in 2015 compared to 52.7% in 2014. During the last 5 years (2013-2017), tetracycline showed the highest resistance rate of 84.7% to *Neisseria gonorrhea* in 2015 while 82% in 2017. The resistance rate of ceftriaxone has slightly reduced to 2.1% in 2017 from 2.2% in 2013. Moreover, the resistance rate of ciprofloxacin and tetracycline has increased by 7% and 5% respectively between 2013 to 2017 (M. o. H. Institute for Medical Research, Malaysia, 2017, 15th September) In addition, the resistance rate for penicillin was 23.5 % in 2010 compared to 61% in 2007 (Lahra, 2012).

## 4. Conclusions

Among all the observed gram-negative bacteria over the years, an increasing trend of antibiotic resistance has revealed the prevalence rate of antibiotic resistance which leads to this study for a review. Thousands of bacterial isolates were examined from different place and locations all over Malaysia. Findings from this preliminary study indicated the prevalence rate of antibiotic-resistant bacteria in Peninsular Malaysia. The outcome of this study can be efficient on infection control measures, and antibiotic stewardship programs should be able to control the spread of the multidrug-resistant bacteria. This could be a proper guideline for appropriate antibiotics to use against gram-negative bacteria. The Most frequent antibiotic resistance profiles were observed by ampicillin, piperacillin, ceftazidime, and tetracycline, etc. regardless in Malaysia. The presence of multidrug resistant bacteria is a significant concern in human health.

In summary, an increasing trend of antibiotic resistance observed over the years. The emergence and increase of antimicrobial resistance are a constant threat and challenge to control infections in Malaysia. The epidemiological infections due to antimicrobial-resistant bacteria have emerged in Malaysia. This study provides a scenario of recent antimicrobial uses all over Malaysia and this will be an aid to reduce acquisition of MDR. Limited infection control measures, lack of antimicrobial control, and travel throughout this region have likely contributed to the increase in multidrug resistant gram negavtive (MDRGN) prevalence. A continuous surveillance programme establishment might help in minimizing such infections. Thus, improving infection control practices, laboratory detection, judicious use of antimicrobial agents, and national surveillance could have an impact on antibiotic resistance as well as MRD spread in Malaysia.

**Acknowledgement**

Authors thankful to Universiti Malaysia Pahang, Malaysia for providing facilities under PGRS-180360 and the authors declare no conflict of interest in the conduct of this study.

**Table 1: The antibiotic resistance rates of *Acinetobacter* species isolates**

| Year | Antibiotics | Highest Resistance % | Source |
|---|---|---|---|
| 2014 | Tetracycline | 87% | Soo-Sum Lean et al. |
| 2012 | Meropenem | 92.5% | Nazmul et al. |
| 2015 | Meropenem | 58.8% | *NSAR |
| 2016 | Meropenem | 60.8% | *NSAR |
| 2016 | Cephalexin | 95.45% | S.Y. Ng et al. |
| 2017 | Imipenem, Ceftazidime, Cefuroxime | - | Selvi et al., |
| | Ceftriaxone | 100% | |
| 2016 | Polymyxin B, Ampicillin/sulbactam, ceftazidime, etc | - | Nathan et al. |
| 2007 | Imipenem, Ceftazidime, Cefuroxime etc | - | Palasubramaniam et al. |
| 2017 | Meropenem | 61.1% | *NSAR |

*NSAR: **National Surveillance of Antimicrobial Resistance, Malaysia**

**Table 2: The highest antibiotic resistance rates of *Klebsiella* spp. among all the clinical isolates**

| Year | Antibiotics | Resistance % | Source |
|---|---|---|---|
| **2017** | **Tetracycline** | **-** | **Low et al.** |
| **2015** | **Ampicillin** | **84.3%** | **Nor et al** |
| **2015** | **Ampicillin** | **100%** | **Farah et al.** |
| | **Aztreonam** | **100%** | |
| | **Cefoperazone** | **100%** | |
| **2015** | **Gentamicin** | **74%** | **Al-Marzooq et al** |
| | **Trimethoprim/sulfamethoxazole** | **90.3%** | |
| **2015** | **Cephalosporin groups** | **-** | **Nurul et al** |

**Table 3: The antibiotic resistance rates of Escherichia coli isolated**

| Year | Antibiotics | Highest Resistance % | Least Resistance % | Source |
|---|---|---|---|---|
| 2009 | Ampicillin | 77 | - | Lim et al |
| - | Amikan | | 2 | - |
| 2013 | Ampicillin | 68.9 | - | *NSAR |
| - | Meropenem | | 0.3 | - |
| 2014 | Ampicillin | 66.20 | - | *NSAR |
| - | Meropenem | | 0.2 | - |
| 2015 | Ampicillin | 67 | - | *NSAR |
| - | Meropenem | | 0.50 | - |
| 2015 | Ampicillin | 67.70 | - | Nor et al |
| - | Gentamicin | | 7.30 | - |
| 2016 | Ceftriaxone | 42.30 | - | Poly et al |
| - | Polymyxin B | | 23.1 | - |
| 2017 | Cefuroxime | 25.9 | | *NSAR |
| | Imipenem | | 0.8 | |

*NSAR: National Surveillance of Antimicrobial Resistance, Malaysia

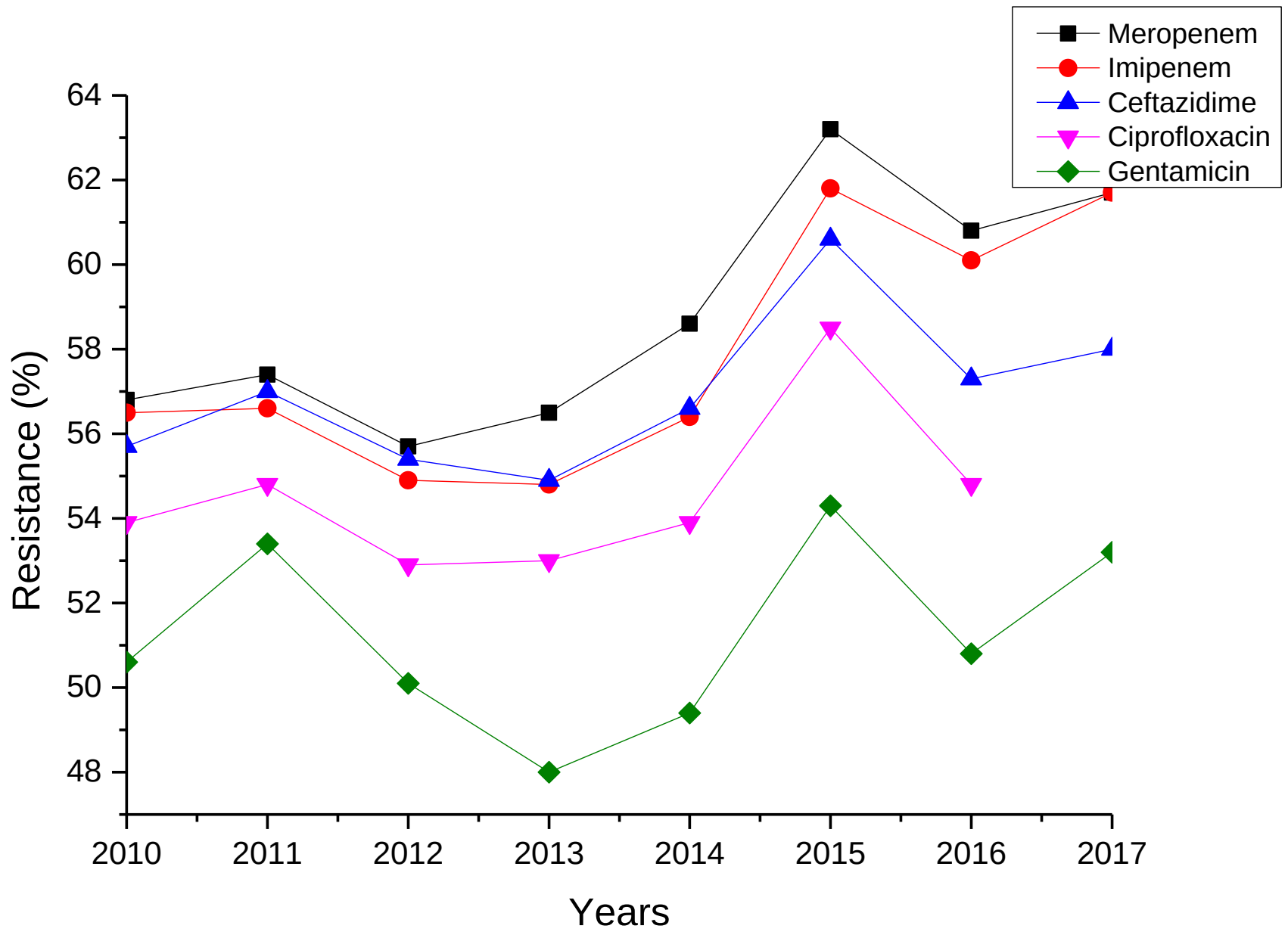

**Figure 1: Antibiotic resistance trends of *Acinetobacter* spp. (2010-2017)**

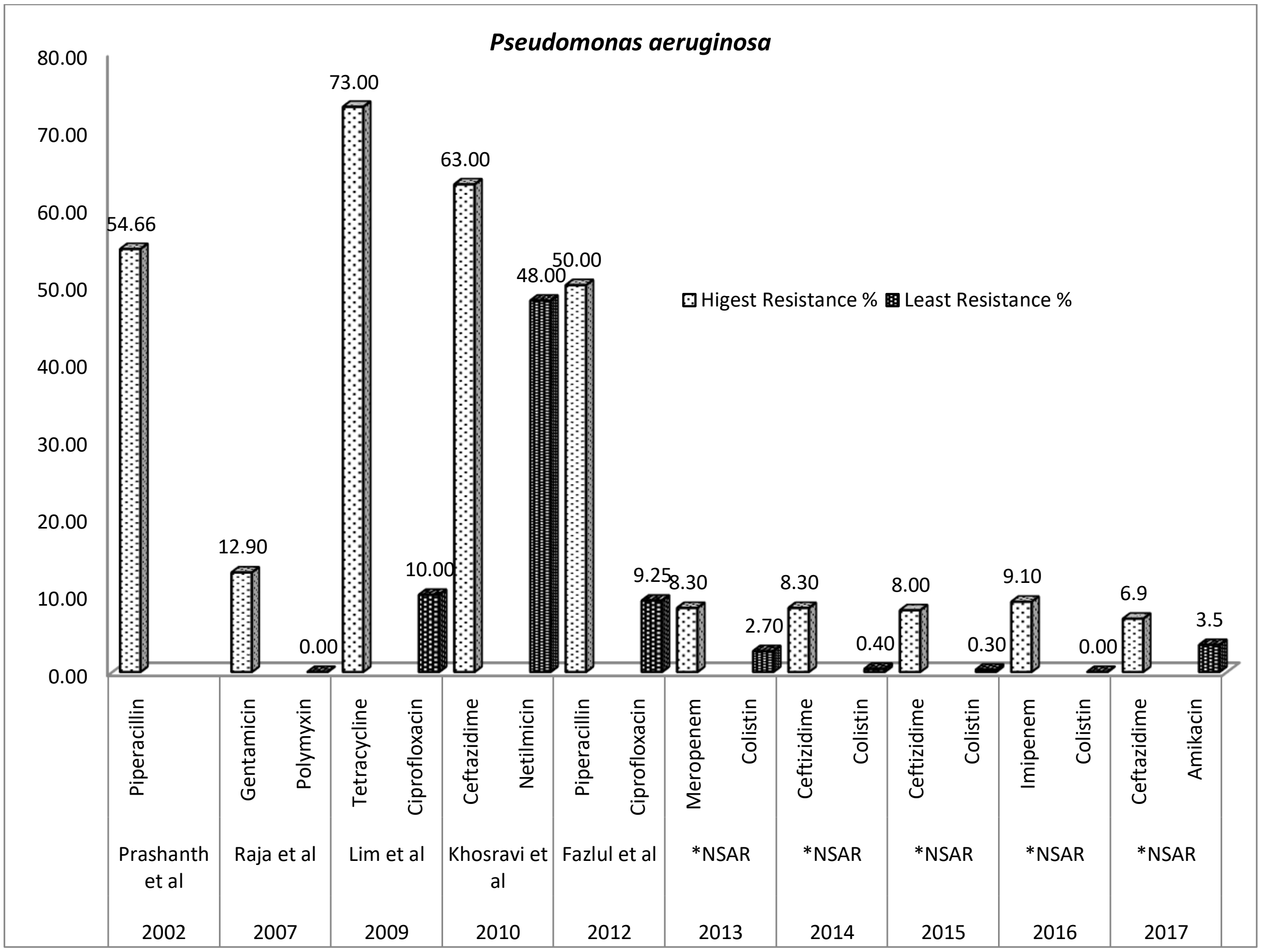

**Figure 2: The antibiotic resistance rates of *Pseudomonas aeruginosa* species isolated *NSAR:** National Surveillance of Antimicrobial Resistance, Malaysia

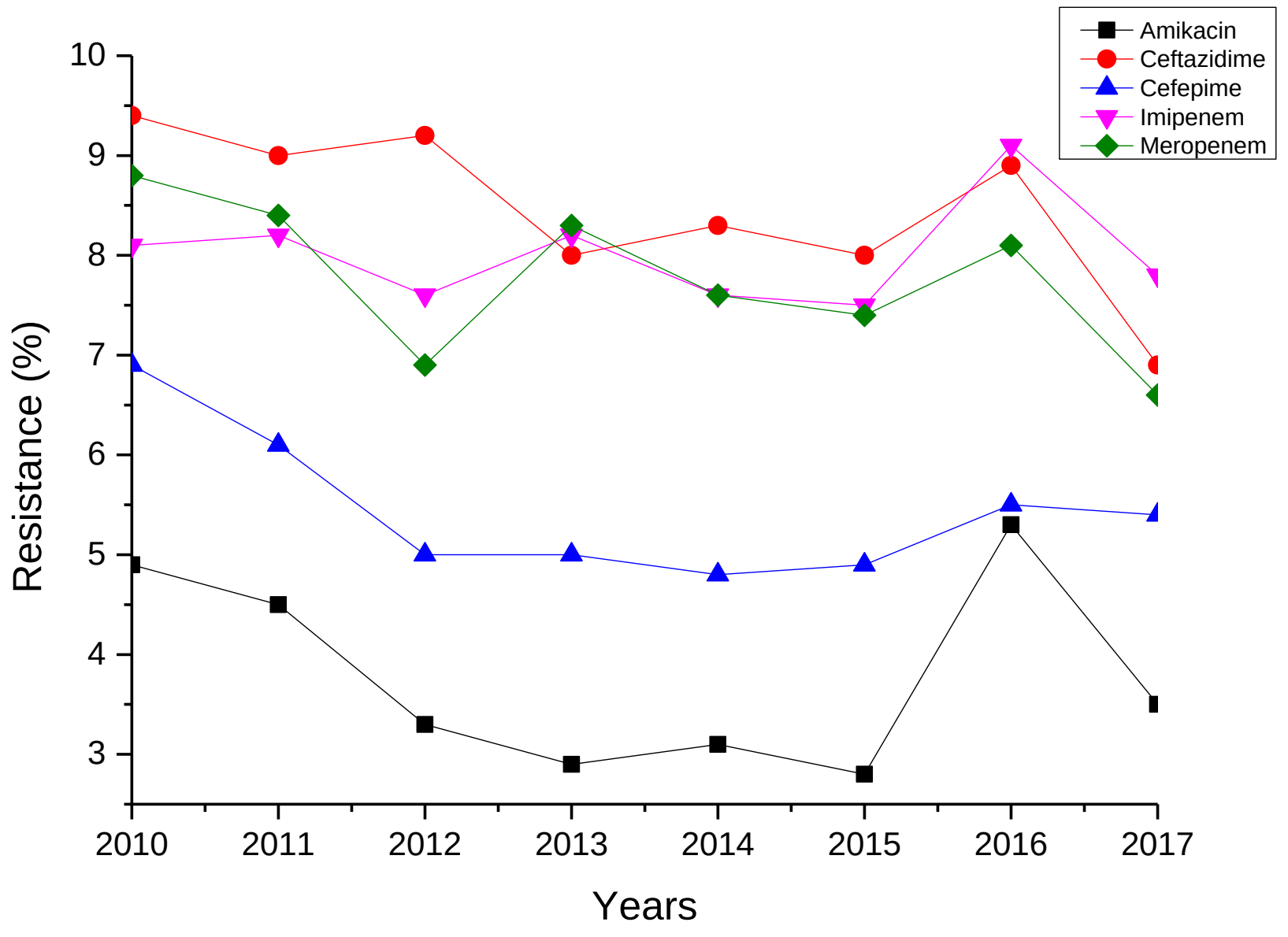

**Figure 3: Antibiotic resistance trends on *Pseudomonas aeruginosa* (2010-2017)**

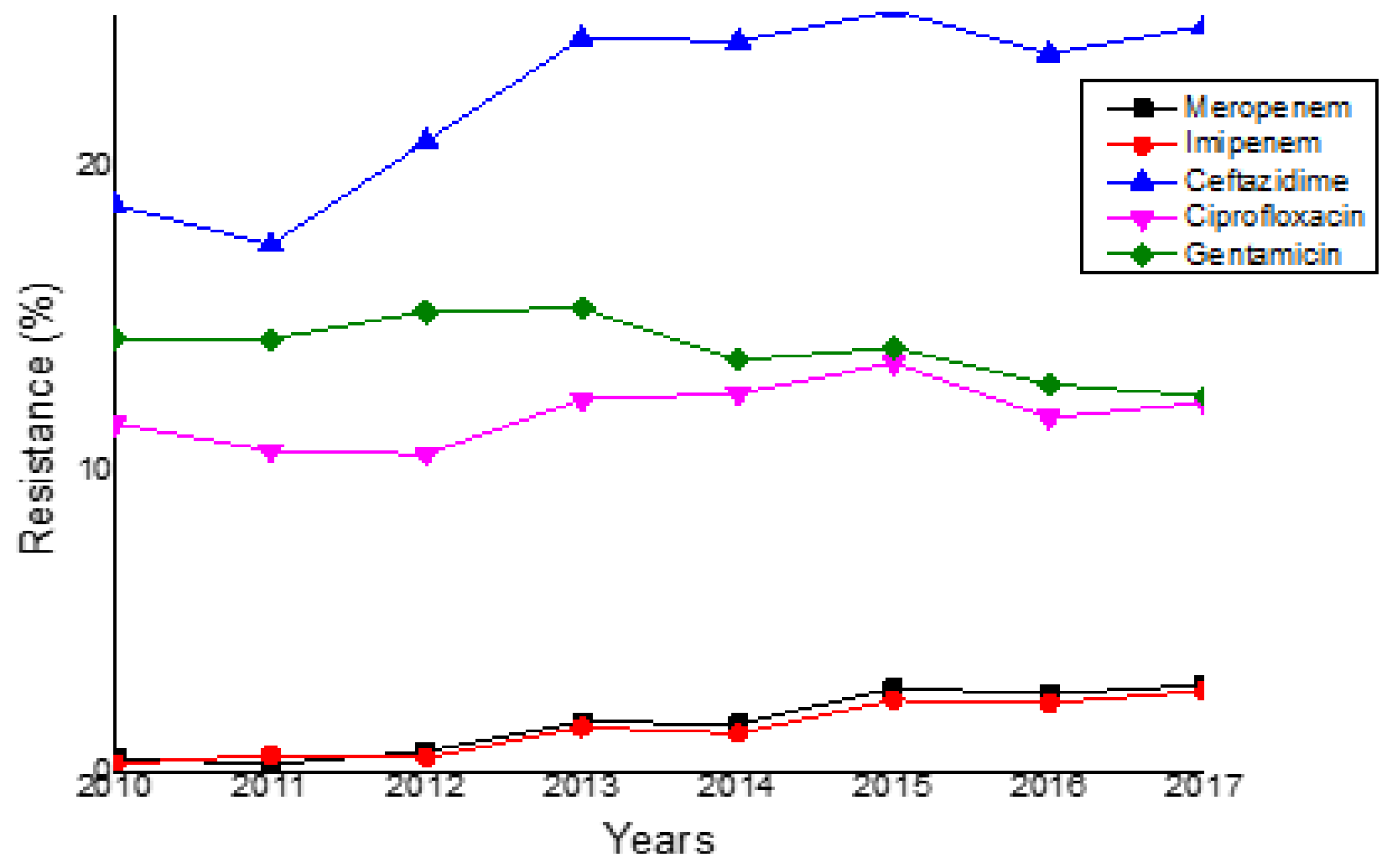

**Figure 4: Antibiotic resistance trends of *Klebsiella* spp. (2010-2017)**

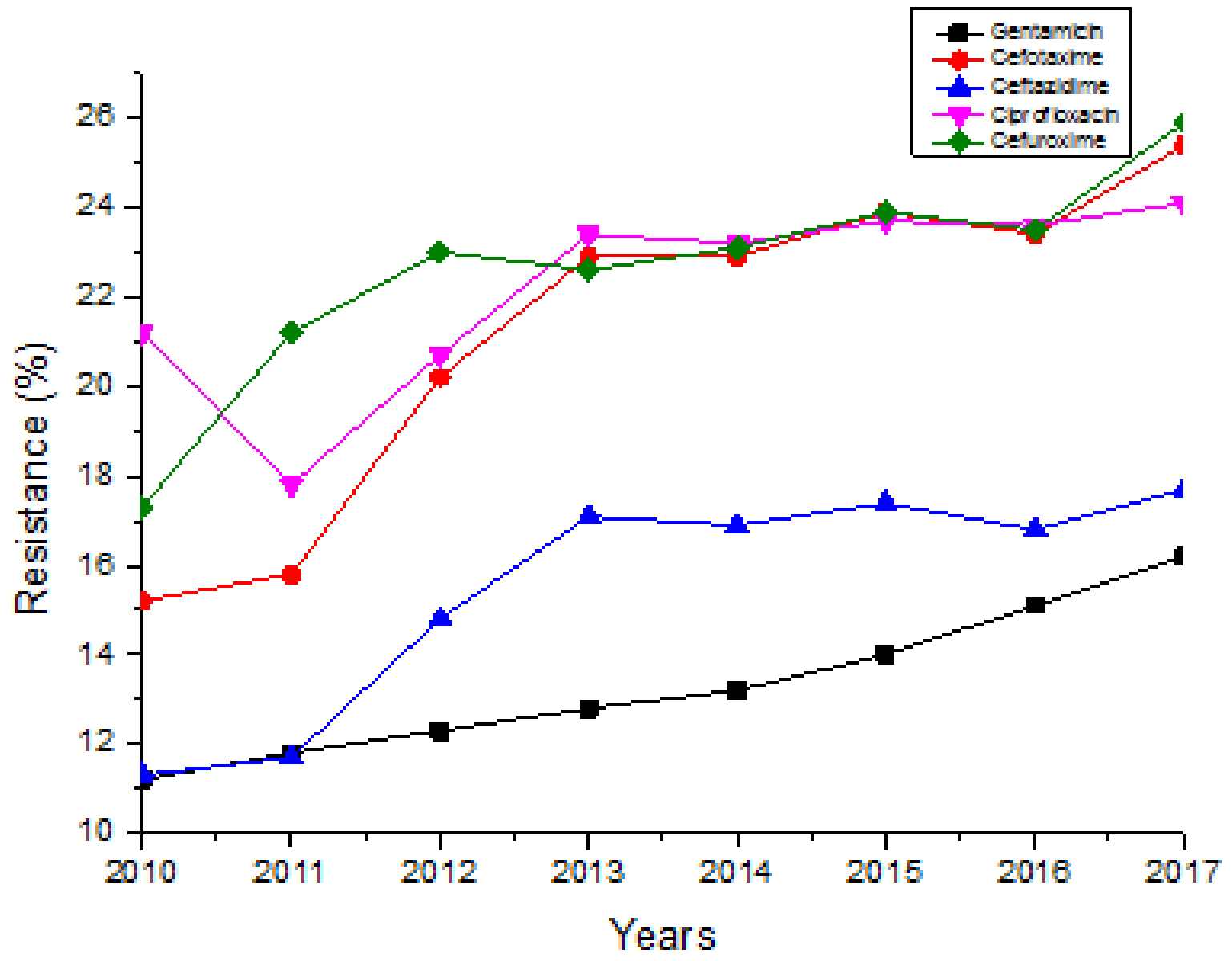

**Figure 5: Antibiotic resistance trends of *E. coli* (2010-2017)**

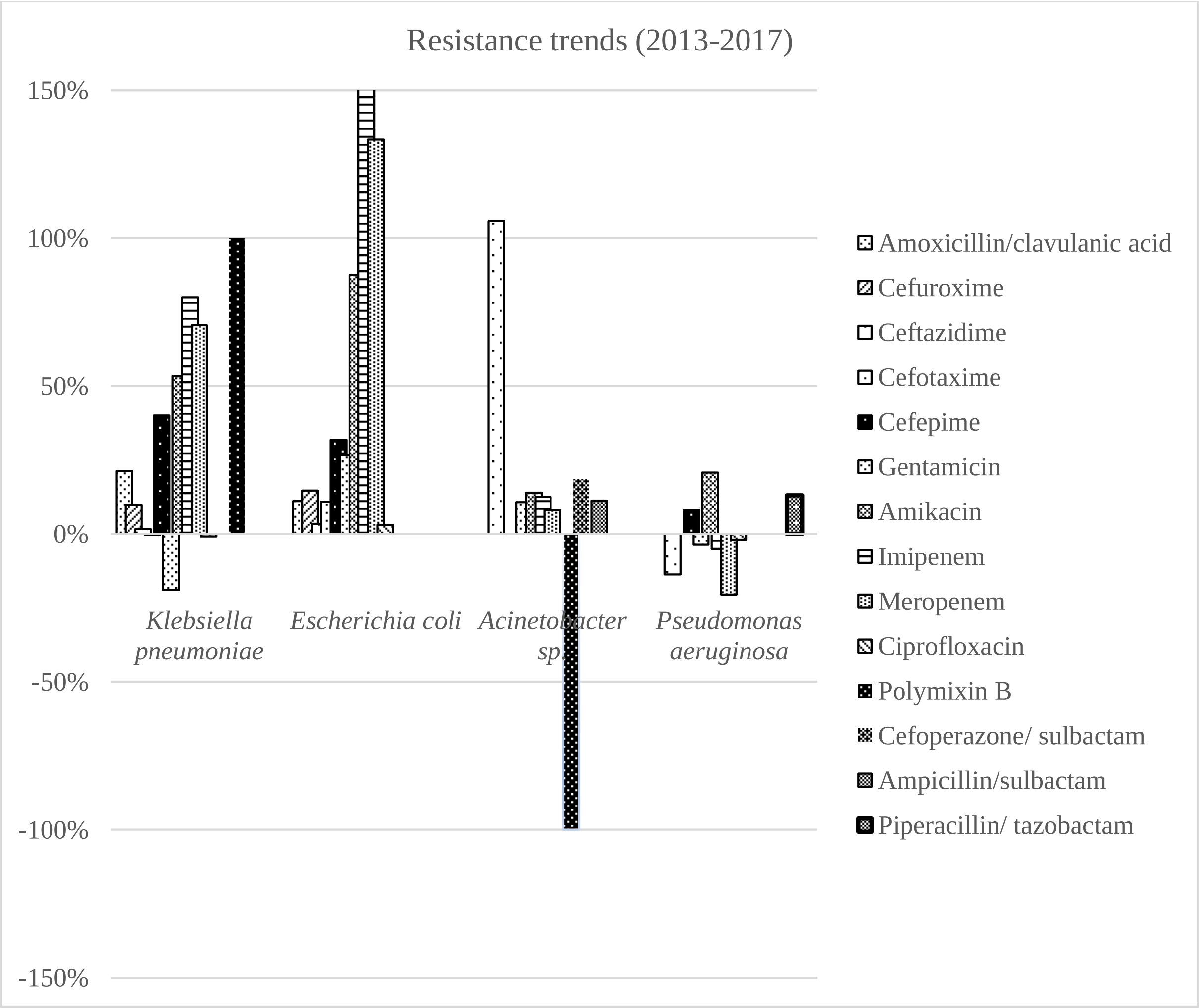

**Figure 6: Antibiotic resistance trends among the most dominant bacteria**